\documentclass[final,3p,10pt,times]{elsarticle}
\usepackage{amsmath,geometry,graphicx,amsbsy,epsfig,latexsym,setspace,wasysym,pifont,mathrsfs} 
\usepackage{lineno}
\usepackage{float}
\usepackage{comment}
\newfloat{widefig}{thp}{lop}
\journal{Physica A}

\begin{document}

\begin{frontmatter}
\title{Finite size scaling study of a two parameter percolation model}
\author{Bappaditya Roy and S. B. Santra} 

\address{Department of Physics, Indian Institute of Technology
  Guwahati, Guwahati-781039, Assam, India.}

\date{\today}

\begin{abstract}
A two parameter percolation model with nucleation and growth of finite
clusters is developed taking the initial seed concentration $\rho$ and
a growth parameter $\mathtt{g}$ as two tunable parameters.
Percolation transition is determined by the final static configuration
of spanning clusters. A finite size scaling theory for such transition
is developed and numerically verified. The scaling functions are found
to depend on both $\mathtt{g}$ and $\rho$. The singularities at the
critical growth probability $\mathtt{g}_c$ of a given $\rho$ are
described by appropriate critical exponents. The values of the
critical exponents are found to be same as that of the original
percolation at all values of $\rho$ at the respective $\mathtt{g}_c$.
The model then belongs to the same universality class of percolation
for the whole range of $\rho$.
\end{abstract}


\end{frontmatter}

\section{Introduction} 
Percolation is one of the most discussed models of statistical physics
of phase transitions that started with the work of Flory in the year
1940 \cite{flory} and proposition of a mathematical model by Broadbent
and Hammersley in 1957 \cite{broad} using the geometrical and
probabilistic concepts. Percolation has found extensive applications
in different branches of science, such as oil recovery from porous
media \cite{king}, epidemic modeling \cite{cardy}, networks
\cite{cohen, Acin}, fracture \cite{roux}, metal-insulator transition
\cite{ball}, ionic transport in glasses and composites \cite{roman},
ground water flow in fractured rocks \cite{sahimi} and many
others. Percolation refers to the formation of long-range
connectedness in a system and is known to be a continuous phase
transition from a disconnected to a fully connected phase at a sharply
defined percolation threshold value \cite{stauffer, ziffp}. Beside the
uncorrelated ordinary percolation (OP), several correlated percolation
models such as bootstrap percolation \cite{bsp}, directed percolation
\cite{dp}, spiral percolation \cite{sp}, directed spiral percolation
\cite{dpsp} are also studied extensively and several non-trivial
features were reported. Percolation theory was also applied to study
random growth process \cite{bunde} such as epidemic spreading, rumor
propagation, etc. A series of non-equilibrium growth models were
proposed in the recent past to demonstrate first order transition that
occurs in an explosive manner \cite{epj223,saberi} after the
introduction of explosive percolation (EP) by Achlioptas {\em et al}
\cite{AP} in contradiction to second order transition in ordinary
percolation. However, most of the EP models are found lacking of
several features of first-order transition such as phase co-existence,
nucleation, etc.  \cite{cep,janssen2016}.

In this paper, a two parameter percolation model (TPPM) with
nucleation and growth of multiple finite clusters simultaneously is
proposed taking the initial seed concentration $\rho$ and a growth
parameter $\mathtt{g}$ as two tunable parameters. As in OP and unlike
growth models, the percolation transition (PT) in this model is
determined by the final static or equilibrium spanning cluster
configurations. For a given $\rho$, a critical growth probability
$\mathtt{g}_c$ is found to exist at which a percolation transition
occurs. It is then intriguing to characterize the properties as well
as the nature of such percolation transitions. It is important to
obtain a phase diagram in the $p-\mathtt{g}$ parameter space that
separates the disconnected phase from the fully-connected phases.

\begin{figure*}
  \centerline{ \hfill\psfig{file=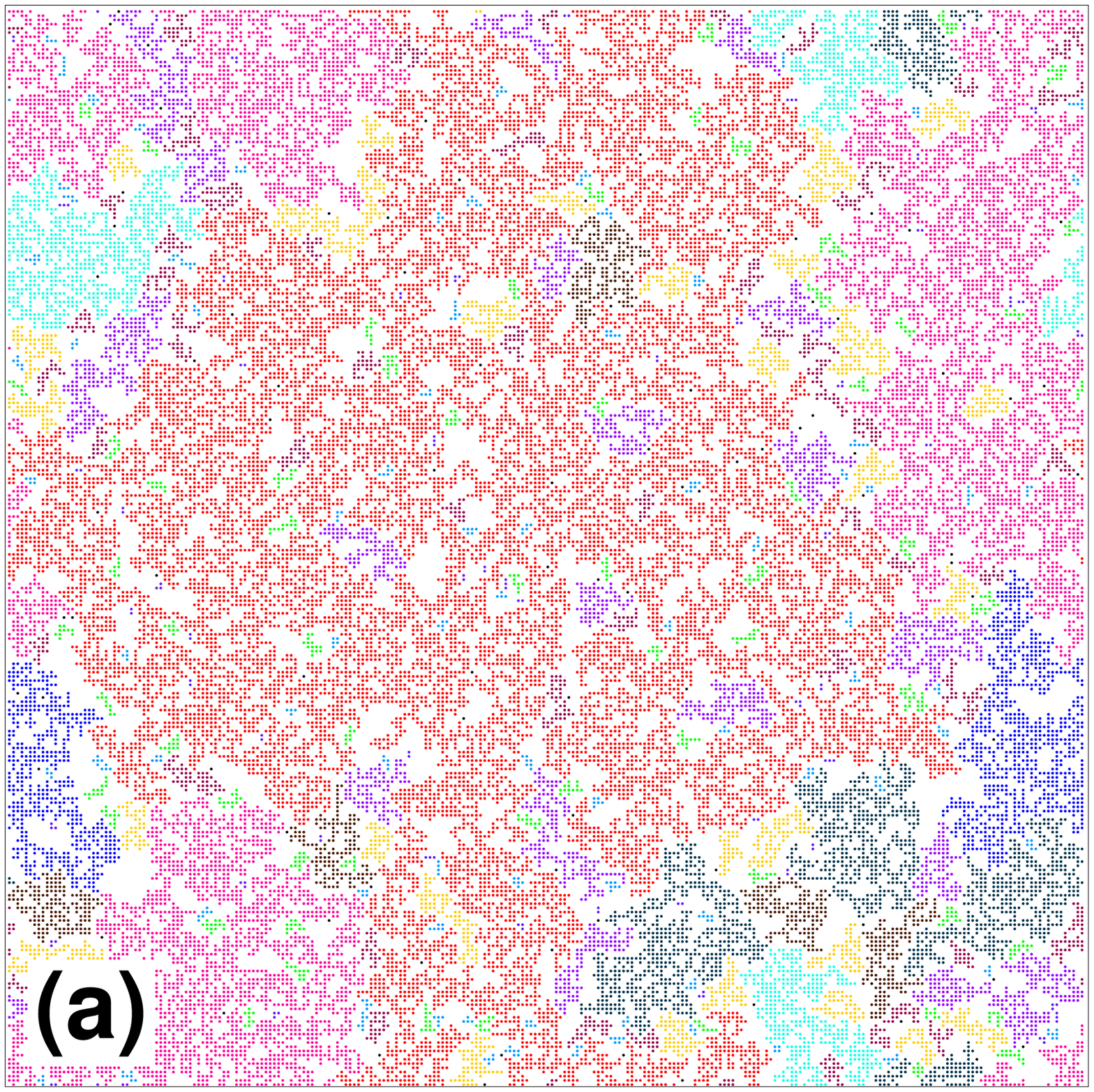, width=0.28\textwidth}
    \hfill\psfig{file=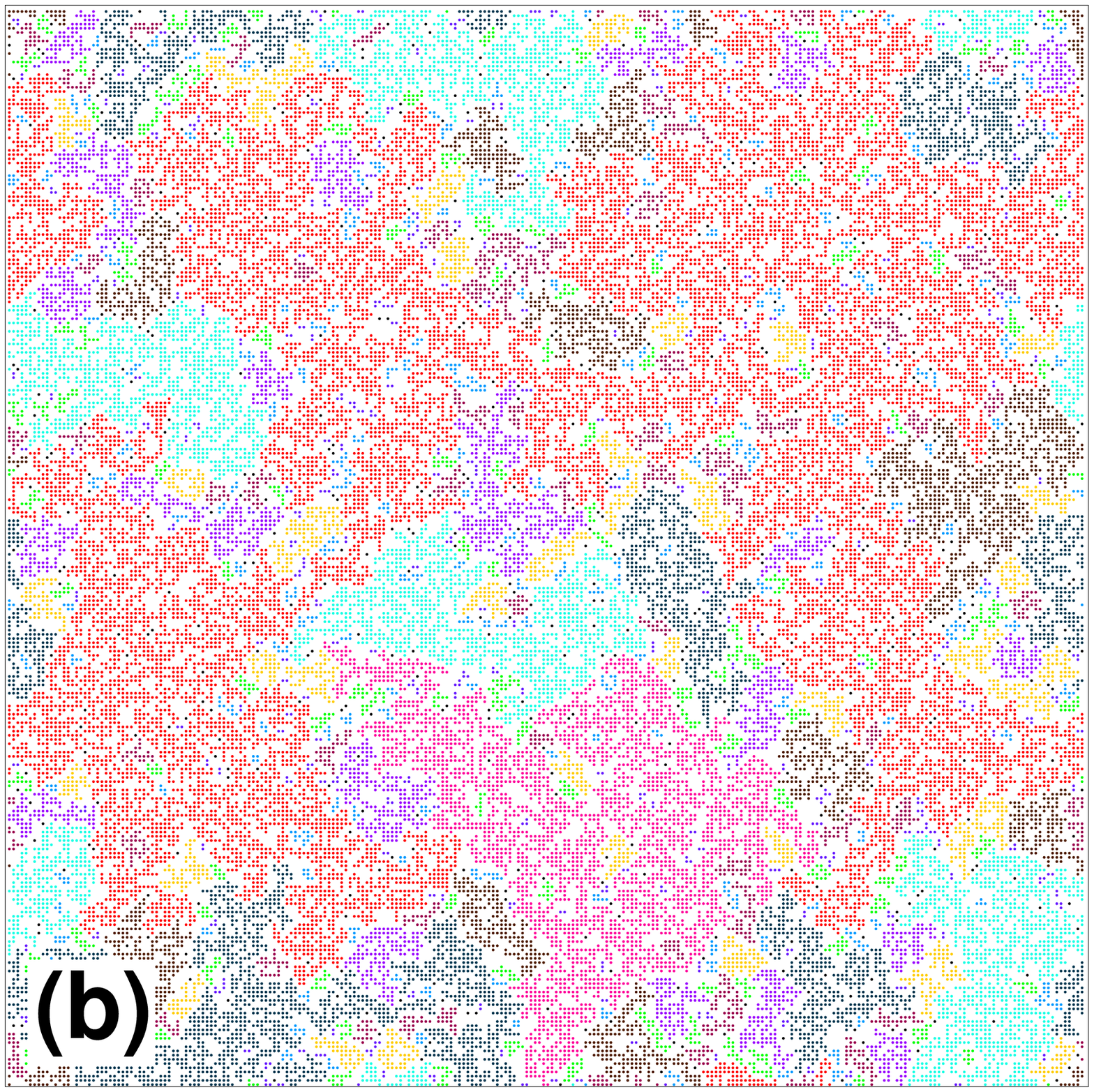, width=0.28\textwidth}\hfill
    \psfig{file=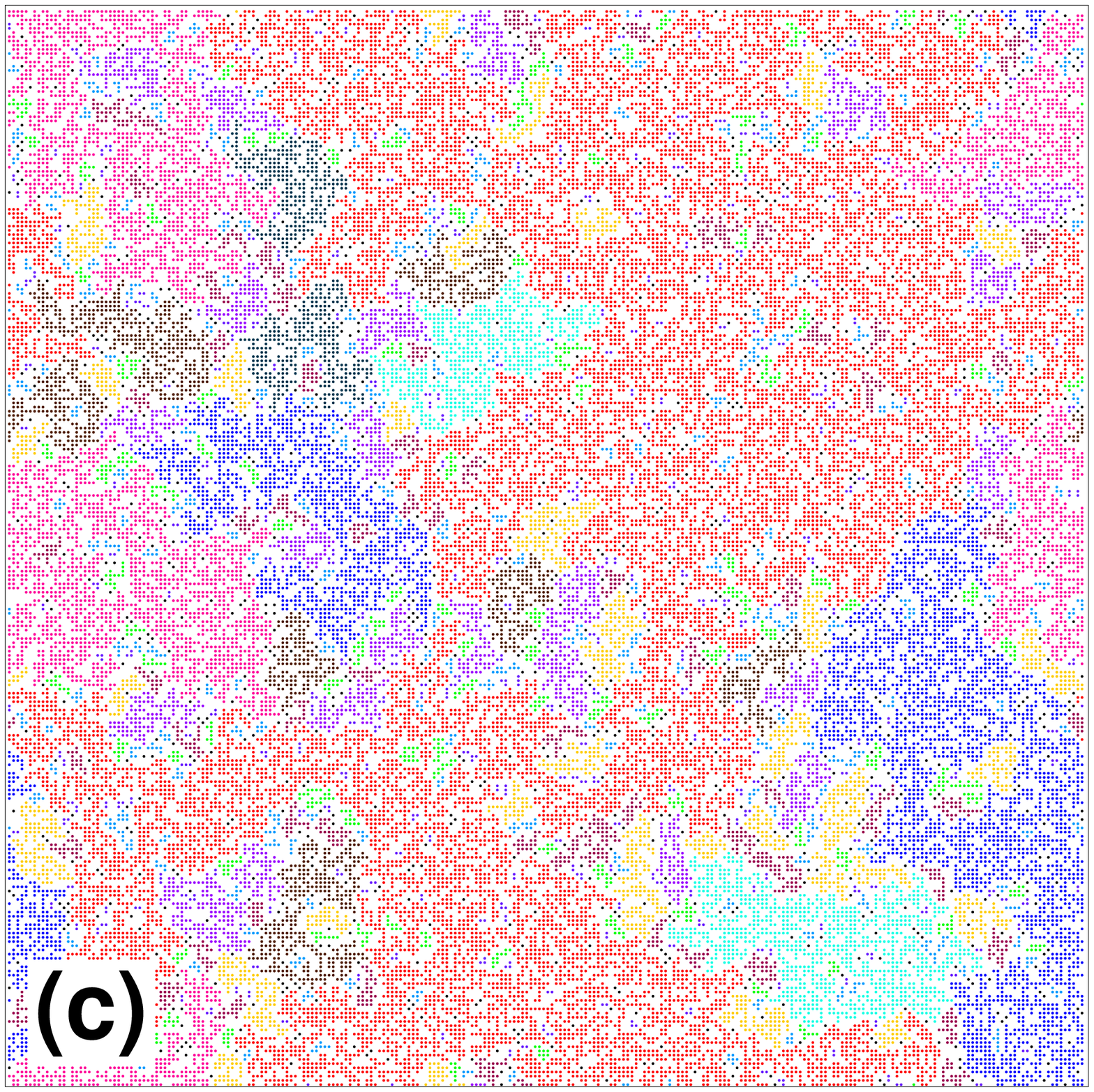, width=0.28\textwidth}\hfill }
\caption{\label{pcbn} Snapshots of cluster configurations at the end
  of the growth process on a $2d$ square system of size $L=256$ with
  initial seed concentration $\rho=0.05$ (a), $0.25$ (b) and $0.50$
  (c) at their respective percolation thresholds
  $\mathtt{g}_c(L)=0.560, 0.440, 0.161$.  Solid line represents the
  lattice boundary. The different colors indicate clusters of
  different sizes. The spanning clusters are shown in red.}
\end{figure*}

\section{Model and simulation} 
In TPPM, the initial configuration is taken as a partially randomly
populated lattice with an initial seed concentration $\rho$ less than
$p_c$, the threshold of OP and the cluster size distribution is
obtained employing Hoshen-Kopelman algorithm \cite{HK}. The process of
nucleation and growth is then implemented by growing all these finite
clusters simultaneously with a constant growth probability
$\mathtt{g}$. Note that there are multiple growth centers instead of a
single one as in Leath's OP \cite{LH}. Once a site is rejected with
probability $(1-\mathtt{g})$, it will remain unoccupied forever, as in
ordinary percolation. During the growth of these clusters, two
clusters may come in contact. Two clusters with occupied perimeter
sites separated by a single lattice spacing are considered to be a
single cluster. The total number of clusters is then reduced by one
and a cluster of larger size is incorporated in the cluster size
distribution. The growth of a cluster stops when there is no empty
site on the cluster perimeter is available to occupy. As the process
stops, the final cluster size distribution function
$n_s(\rho,\mathtt{g})$ is obtained. The model has two limiting
situations. One is $\rho=p_c$, the critical site occupation
probability of percolation and $\mathtt{g}=0$. The other one is
$\rho=1/L^2$, a single seed and $\mathtt{g}=p_c$. Both the situations
correspond to ordinary percolation problem. The present model can be
considered as a generalized multiple cluster growth model of
percolation. In the following, taking intermediate values of $\rho$
and varying the growth probability $\mathtt{g}$ transitions from
disconnected phase to fully connected phase are studied.

An extensive computer simulation has been performed on two dimensional
($2d$) square lattices of size $L\times L$ varying $L$ from $2^7$ to
$2^{11}$. For a given $L$, initial seed concentration $\rho$ is varied
between $1/L^2$ and $0.59$. All finite clusters are identified and
grown by occupying the empty nearest neighbours (NN) of the perimeter
(both internal and external) sites of these clusters with probability
$\mathtt{g}\in [0,1]$. In growing the clusters periodic boundary
conditions (PBC) are applied in both the directions. Since the
clusters are grown applying PBC, the horizontal and vertical
extensions of the largest cluster is kept stored. If either the
horizontal or the vertical extension of the largest cluster is found
to be $\ge L$, the system size, it is identified as a spanning
cluster. The percolation thresholds $\mathtt{g}_c$, the probability at
which a spanning cluster appears for the first time in the system as
$\mathtt{g}$ is increasing to a critical value, are estimated for each
$\rho$ on a given $L$.  Data are averaged over $10^5$ to $10^6$
ensembles for each parameter set. Snapshots of the system morphology
at the end of the growth process on a square lattice of size $L=256$
with initial seed concentrations $\rho=0.05, 0.25$ and $0.50$ are
shown in Fig.\ref{pcbn} at their respective thresholds
$\mathtt{g}_c(L)=0.560, 0.440, 0.161$. In these snapshots, different
colors indicate clusters of different sizes. White space corresponds
to inaccessible lattice sites. It can be noticed that at the high
$\rho$ inaccessible area is less than that at small $\rho$ at their
respective thresholds. The spanning cluster is shown in red.
Interestingly, irrespective of the values of $\rho$, it seems cluster
of all possible sizes appear at their respective percolation
thresholds indicating continuous phase transition for all values of
$\rho$.

\section{Percolation threshold, Critical exponents and Scaling}
A scaling theory for TPPM is developed following the techniques of
ordinary percolation. In the present model, one starts with an initial
seed concentration $\rho$ and the empty sites around the clusters
formed by the initial seeds are grown with probability
$\mathtt{g}$. The area fraction $p$, number of occupied sites per
lattice site, at the end of the growth process is expected to be
\begin{equation}
\label{prg}
p=\rho+\mathtt{g}(1-\rho)
\end{equation}
if all the remaining empty sites are called for occupation. Since we
followed cluster growth process to populate the lattice, it may not
always be possible to call all the empty sites except for
$\mathtt{g}=1$. As a result, a small area fraction remain inaccessible
at the end of the growth process for smaller values of
$\rho$. However, for a given $\rho$, the difference in area fractions
$p-p_c$ corresponding to the growth probability $\mathtt{g}$ and that
with $\mathtt{g}_c$, the critical threshold, will always be
proportional to $(\mathtt{g}-\mathtt{g}_c)(1-\rho)$ in the critical
regime. Hence the scaling form of the cluster size distribution and
that of all other related geometrical quantities in TPPM can be
obtained in terms of $\mathtt{g}$ and $\rho$ by substituting $p-p_c$
by $(\mathtt{g}-\mathtt{g}_c)(1-\rho)$ in the cluster size
distribution of OP as
\begin{equation}
\label{nsg}
n_s(\rho,g)=s^{-\tau}\widetilde{n}_{s}
[(\mathtt{g}-\mathtt{g}_c)(1-\rho)s^\sigma]
\end{equation}
where $f$ is a new scaling function and $\tau,\sigma$ are new scaling
exponents. The scaling form of different geometrical quantities in
terms of $\rho$ and $\mathtt{g}$ can be derived from the above cluster
size distribution $n_s(\rho,\mathtt{g})$ as per their original
definitions in terms of $n_s(p)$.

\subsection{Percolation threshold}
 \begin{figure*}[t]
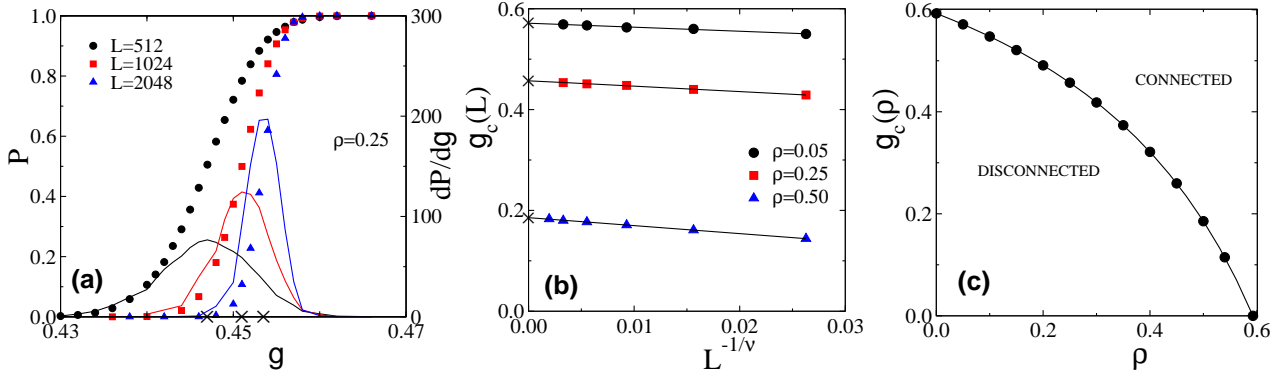

 \centerline{\hfill
   \psfig{file=santra_fig2a.eps,width=0.36\textwidth} \hfill
   \psfig{file=santra_fig2b.eps,width=0.32\textwidth}
   \psfig{file=santra_fig2c.eps,width=0.32\textwidth}\hfill}
 \caption{\label{pc} (a) Plot of spanning probability $P$ and its
   derivative $dP/d\mathtt{g}$ against $\mathtt{g}$ for different
   system sizes $L$ taking $\rho=0.25$. The symbols are: circle for
   $L=512$, squares for $L=1024$ and triangles for $L=2048$ and the
   derivatives are represented by a black line, a red lines and a blue
   line respectively. The values of $\mathtt{g}$ corresponding to the
   maxima of the derivatives indicate $\mathtt{g}_c(L)$ and marked by
   crosses. (b) Plot of $\mathtt{g}_c(L)$ versus $L^{-1/\nu}$ for
   $\rho=0.05 (\bigcirc), 0.25 (\square)$ and $0.50 (\triangle)$. The
   best straight line fit is found for $\nu=4/3$. From the intercepts,
   $\mathtt{g}_c(\rho)$ are obtained as $0.5713, 0.4570, 0.1855$
   respectively. (c) Plot of $\mathtt{g}_c(\rho)$ against $\rho$. The
   circles represent the estimated threshold and the line represents
   the analytic value obtained from Eq.\ref{pheq}.}
 \end{figure*}

Percolation threshold is identified as the critical growth probability
$\mathtt{g}_c$ for a given $\rho$ at which for the first time a
spanning cluster connecting the opposite sides of the lattice appears
in the system. In order to calculate $\mathtt{g}_c$ of a given $\rho$
and system size $L$, the probability $P(\rho,\mathtt{g},L)$ to get a
spanning cluster is defined as
\begin{equation}
\label{pspd}
P(\rho,\mathtt{g},L)=\frac{N_{sp}(\rho,\mathtt{g},L)}{N_{tot}}=f[(\mathtt{g}-\mathtt{g}_c(\rho))(1-\rho)L^{1/\nu}]
\end{equation}
out of $N_{tot}$ ensembles for a system size $L\ll\xi$, the
correlation length, $\nu$ is the correlation length exponent. In the
limit $L\rightarrow\infty$, $P(\rho,\mathtt{g},L)$ is expected to be a
theta function and its derivative with respect to $\mathtt{g}$ would
be a delta function at $\mathtt{g}=\mathtt{g}_c(\rho)$. Therefore, for
a given $\rho$, the value of $\mathtt{g}_c(\rho,L)$ at which a
spanning cluster appears for the first time is taken as the mean
values of the distribution $dP/d\mathtt{g}$ as
\begin{equation}
\label{gc}
\mathtt{g}_c(\rho,L)=\int_0^1
\mathtt{g}\frac{dP}{d\mathtt{g}}d\mathtt{g}
=\mathtt{g}_c(\rho)+C\frac{L^{-1/\nu}}{1-\rho},
\end{equation}
where $C=\int_{-\infty}^{+\infty}zf'(z)dz$, for
$z=(\mathtt{g}-\mathtt{g}_c(\rho))(1-\rho)L^{1/\nu}$
\cite{santraroy}. In Fig.\ref{pc}(a), for $\rho=0.25$ the probability
of having a spanning cluster $P(\rho,\mathtt{g},L)$ is plotted against
$\mathtt{g}$ for three different values of $L$. Their derivatives are
shown by continuous lines in the same figure in same color of the
symbols for a given $L$. The value of $\mathtt{g}_c(\rho,L)$ is
identified as the value of $\mathtt{g}$ corresponding to the maximum
of the derivatives and marked by crosses on the $\mathtt{g}$-axis. In
Fig.\ref{pc}(b), $\mathtt{g}_c(\rho,L)$ are plotted against
$L^{-1/\nu}$ taking $\nu=4/3$ as that of percolation for three
different values of $\rho$. It has been verified that the best
straight line was found for $\nu=4/3$. The percolation threshold
$\mathtt{g}_c(\rho)$ for infinite system size is then obtained from
the intercepts with the $y$-axis. Since the other geometrical
properties are evaluated for selective $\rho$ values, the thresholds
$\mathtt{g}_c(\rho)$ are also obtained for several values of $\rho$. A
phase diagram in the $\rho-\mathtt{g}$ parameter space is constructed
by plotting the values of $\mathtt{g}_c(\rho)$ against $\rho$ in
Fig.\ref{pc}(c). The points constitute a phase line that separates the
phase space into percolating and non-percolating regions. It is
interesting to note that line connecting the data points satisfies the
following equation
\begin{equation}
\label{pheq}
\rho+\mathtt{g}_c(1-\rho)=p_c({\rm OP})
\end{equation}
where $p_c(\approx 0.592746)$ is the percolation threshold of OP. The
line of continuous phase transitions terminates at two trivial
critical points corresponding to ordinary percolation.

\begin{figure*}[h]
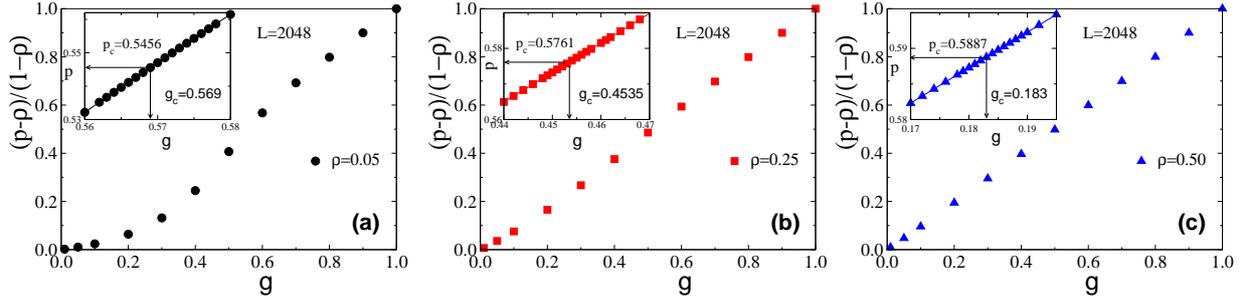

 \centerline{\hfill \psfig{file=santra_fig3a.eps,width=0.32\textwidth}
   \hfill \psfig{file=santra_fig3b.eps,width=0.32\textwidth}
   \psfig{file=santra_fig3c.eps,width=0.32\textwidth}\hfill}
 \caption{\label{pg} (a) Plot of $(p-\rho)/(1-\rho)$ against
   $\mathtt{g}$ for $\rho=0.05$(a), $0.25$(b) and $0.50$(c) and for
   system sizes $L=2048$. In the inset of figures, the area fraction
   $p$ is plotted against $\mathtt{g}$ within the transition
   region. The thresholds value of $\mathtt{g}=\mathtt{g}_c$ and
   corresponding $p=p_c$ for $L=2048$ is marked by arrow for all
   $\rho$ values.}
 \end{figure*}

Before proceeding further, the critical regime is verified by plotting
the variation of $(p-\rho)/(1-\rho)$ against the growth probability
$\mathtt{g}$ in Fig.\ref{pg} for $\rho=0.05$(a), $0.25$(b) and
$0.50$(c) and for the system size $L=2048$.  It is found to be linear
with $\mathtt{g}$ for higher $\rho$ values. Whereas non-linearity
arises for smaller values of $\rho$. In the inset, measured $p$ is
plotted against $\mathtt{g}$ and found proportional for all values of
$\rho$ within the critical region. The thresholds growth probability
$\mathtt{g}_c$ and the corresponding area fraction $p_c$ for $L=2048$
are shown by arrows for all values of $\rho$. The nature of transition
at the intermediate values of $\rho$ will be determined now.


\begin{figure*}[t]
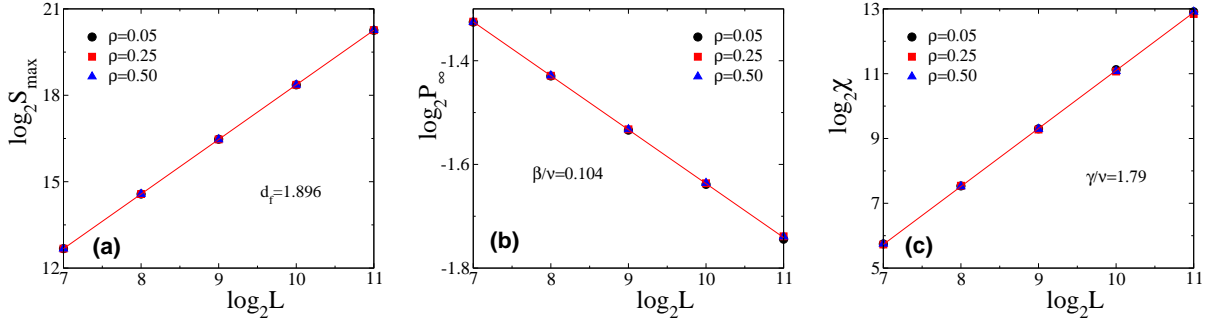

  \centerline{\hfill\psfig{file=santra_fig4a.eps, width=0.3\textwidth}
    \hfill\psfig{file=santra_fig4b.eps, width=0.3\textwidth}
    \hfill\psfig{file=santra_fig4c.eps, width=0.3\textwidth}\hfill}
\caption{\label{dfbetagamma} Plot of $S_{max}$, $\chi'^2$,
  $\chi_{\infty}$ and $P_{\infty}$ against system sizes $L$ at their
  respective thresholds $\mathtt{g}_c(\rho)$ for three different
  values of $\rho=0.05(\bigcirc)$ (black), $0.25(\square)$(red) and
  $0.50(\triangle)$(blue). The fractal dimension $d_f$= 1.896 in (a),
  $\beta/\nu=0.104$ in (b) and $\gamma/\nu=1.79$ in (c) are estimated
  from the slope of the straight line and found to be independent of
  $\rho$.}
\end{figure*}

\subsection{Critical exponents}
Following the cluster size distribution is given in Eq.\ref{nsg}, the
order parameter $P_\infty$ and the average cluster size $\chi$ can be
defined in terms of $\rho$ and $\mathtt{g}$ as
\begin{equation}
\label{pinfeq}
P_{\infty} =\frac{S_{max}}{L^2}= \rho+\mathtt{g}(1-\rho)
-\sideset{}{'}\sum_ssn_s(\rho,\mathtt{g})
\end{equation} 
and  
\begin{equation}
\label{chi}
\chi=\sideset{}{'}\sum_ss^2n_s(\rho,\mathtt{g})
\end{equation}
where the primed sum indicates that the spanning cluster is excluded.
The percolation spanning cluster is a random object with all possible
holes in it and is expected to be fractal. For system size $L$, the
size of the spanning cluster $S_{max}$, at the percolation threshold
varies with the system size $L$ as
\begin{equation}
\label{fd}
\langle S_{max}\rangle \approx L^{d_f}
\end{equation}
where $d_f$ is the fractal dimension of the spanning
cluster. Following scaling theory of OP, the scaling behavior of
$P_\infty$ and $\chi$ are expected to be
\begin{equation}
\label{pinf1}
P_{\infty} \sim [(\mathtt{g}-\mathtt{g}_c(\rho))(1-\rho)]^\beta \ \ \ {\rm and} \ \ \ 
\chi \sim [(\mathtt{g}-\mathtt{g}_c(\rho))(1-\rho)]^{-\gamma}
\end{equation}
where $\beta=(\tau-2)/\sigma$ and $\gamma=(3-\tau)/\sigma$. Presuming
that the connectivity (correlation) length $\xi \sim
[(\mathtt{g}-\mathtt{g}_c(\rho))(1-\rho)]^{-\nu}$, one could also establish
that $d_f=d-\beta/\nu$. However, the critical exponents measured are
very often found to be limited by the finite system size $L$. A system
is said to be finite if its size $L$ is less than the connectivity
length $\xi$. If a quantity $Q$ is predicted to scale as
$|(\mathtt{g}-\mathtt{g}_c(\rho))(1-\rho)|^{-q}$ for the system size
$L\gg\xi$, then the scaling form of $Q$ for the system size $L\ll\xi$
is expected to be
\begin{equation}
\label{fsc1}
Q(\rho,\mathtt{g},L)=L^{q/\nu}
\widetilde{Q}[(\mathtt{g}-\mathtt{g}_c(\rho))(1-\rho)L^{1/\nu}]
\end{equation}
where $q$ is an exponent $\widetilde{Q}$ is a scaling function. The
finite size scaling form of $P_\infty$ and the average cluster size
$\chi$ are then given by
\begin{equation}
\label{fsc2}
P_\infty(\rho,\mathtt{g},L)=L^{-\beta/\nu}
\widetilde{P}_{\infty}[(\mathtt{g}-\mathtt{g}_c(\rho))(1-\rho)L^{1/\nu}]
\ \ \ {\rm and} \ \ \ \chi(\rho,\mathtt{g},L)=L^{\gamma/\nu}
\widetilde{\chi}[(\mathtt{g}-\mathtt{g}_c(\rho))(1-\rho)L^{1/\nu}].
\end{equation}

The values of $S_{max}$, $P_{\infty}$ and $\chi$ are estimated at
$\mathtt{g}=\mathtt{g}_c(\rho)$ for several systems sizes $L$ as well
as for different values of $\rho$. At $\mathtt{g}=\mathtt{g}_c(\rho)$,
the functions $\widetilde{P}_{\infty}$ and $\widetilde{\chi}$ are
expected to be constants. The values of $S_{max}$, $P_{\infty}$ and
$\chi$ are plotted against $L$ in double logarithmic scales in
Fig.\ref{dfbetagamma} (a), (b) and (c) respectively. It can be seen
that they follow the respective scaling behaviors. The values of
$d_f$, $\beta/\nu$ and $\gamma/\nu$ are estimated by linear least
square fit through the data points. It is found that $d_f=1.896 \pm
0.001$, $\beta/\nu=0.104$ and $\gamma/\nu=1.79$. Though the values of
critical exponents remain same as previously reported
\cite{santraroy}, the precise measurements lead to slight
modifications in the magnitude of the geometrical quantities. The
values of $d_f$ and ratios of the exponents as that of OP and hence
the phase transition are continuous. It is also important to note that
the absolute values of these quantities are independent of the initial
seed concentration $\rho$. This means that the area fraction given in
terms of $\rho$ and $\mathtt{g}$ in Eq.\ref{prg} holds correctly at
the percolation threshold $\mathtt{g}_c(\rho)$ and the spanning
clusters of the same size for different $\rho$ are produced. It could
also be noted here that in the touch and stop model \cite{arg}, for
low concentration of initial seed the final area fraction was found to
be same. The scaling relation $d_f=2-\beta/\nu$ is satisfied here
within error bars because all critical exponents are as that of
percolation.

\subsection{Order parameter and its fluctuation}
Following the formalism of analyzing thermal critical phenomena
\cite{binder, bruce}, the distribution of $P_\infty$ is taken as
\begin{equation}
\label{pinfd}
P(P_{\infty})=L^{\beta/\nu} \widetilde{P}[P_{\infty}L^{\beta/\nu}]
\end{equation}
where $\widetilde{P}$ is a universal scaling function. Such a
distribution function of $P_\infty$ is also used in the context of PT
recently \cite{cep}. With such scaling form of $P_\infty$
distribution, one could easily show that $\langle P_{\infty}^2\rangle$
as well as $\langle P_{\infty}\rangle^2$ scale as $\sim
L^{-2\beta/\nu}$. The susceptibility is defined in terms of the
fluctuation in $P_\infty$ as
\begin{equation}
\label{chiinf}
\chi_{\infty}= \frac{1}{L^2}[\langle S_{max}^2\rangle - \langle
  S_{max}\rangle^2].
\end{equation}
Following the hyper-scaling relation $d\nu=\gamma+2\beta$, the FSS
form of $\chi_{\infty}$ is obtained as
\begin{equation}
\label{chiinffss}
\chi_{\infty}= L^{\gamma/\nu}
\widetilde{\chi}_\infty[(\mathtt{g}-\mathtt{g}_{c}(\rho))(1-\rho)L^{1/\nu}]
\end{equation}
where $\widetilde{\chi}_\infty$ is a scaling function. The finite size
scaling form of the order parameter $P_\infty(\rho,\mathtt{g},L)$ and its
fluctuation $\chi_{\infty}$ are now verified at different values of
$\rho$.
\begin{figure*}[t]
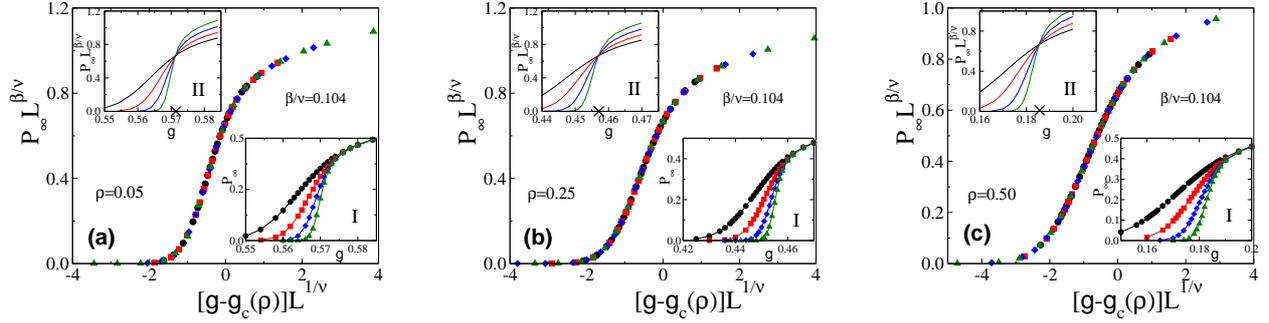

  \centerline{\psfig{file=santra_fig5a.eps, width=0.3\textwidth}
    \hfill\psfig{file=santra_fig5b.eps, width=0.3\textwidth}
  \hfill\psfig{file=santra_fig5c.eps, width=0.3\textwidth}}
\caption{\label{pinf} Plot of $P_{\infty}L^{\beta/\nu}$ against
  $[\mathtt{g}-\mathtt{g}_c(\rho)]L^{1/\nu}$ for $\rho=0.05$(a),
  $0.25$(b) and $0.50$(c) and for $L=256(\bigcirc)$ (black),
  $512(\square)$ (red), $1024(\Diamond)$ (blue) and
  $2048(\triangle)$(green). A good data collapse is obtained for the
  estimated critical exponents $\beta/\nu=0.104$. In inset-I,
  $P_{\infty}$ $vs$ $\mathtt{g}$ and in inset-II,
  $P_{\infty}L^{\beta/\nu}$ $vs$ $\mathtt{g}$.}
\end{figure*}

In Fig.\ref{pinf}, the variation of $p_\infty$ is studied for
$\rho=0.05$ (a), $\rho=0.25$ (b) and $\rho=0.50$ (c). In the inset-I
of Fig.\ref{pinf}, $P_\infty$ is plotted against the growth parameter
$\mathtt{g}$ for different system sizes $L$ at each
$\rho$. Irrespective of the value of $\rho$, the transition become
sharper and sharper as $L\rightarrow\infty$ as expected. In the
inset-II of Fig.\ref{pinf}, the scaled order parameter $P_\infty
L^{\beta/\nu}$ is plotted against the growth parameter $\mathtt{g}$
for the same system sizes. A precise crossing point at a particular
$\mathtt{g}$ is observed for the scaled order parameter of different
fixed values of $L$ for a given $\rho$. These crossing points are
verified to the critical thresholds $\mathtt{g}_c$ of the growth
parameter for corresponding values of $\rho$. Finally, the scaled
order parameter $P_\infty L^{\beta/\nu}$ is plotted against the scaled
variable $[\mathtt{g}-\mathtt{g}_c(\rho)]L^{1/\nu}$ for different
system sizes $L$ at each $\rho$. For each $\rho$, the values of
$\beta/\nu$ and $1/\nu$ are taken as that of OP. A good data collapse
is observed for all values of $L$ at every value of $\rho$. The
distribution of order parameter is found to be a single humped
distribution at all values of $\rho$ as in continuous phase
transition.

\begin{figure*}[h]
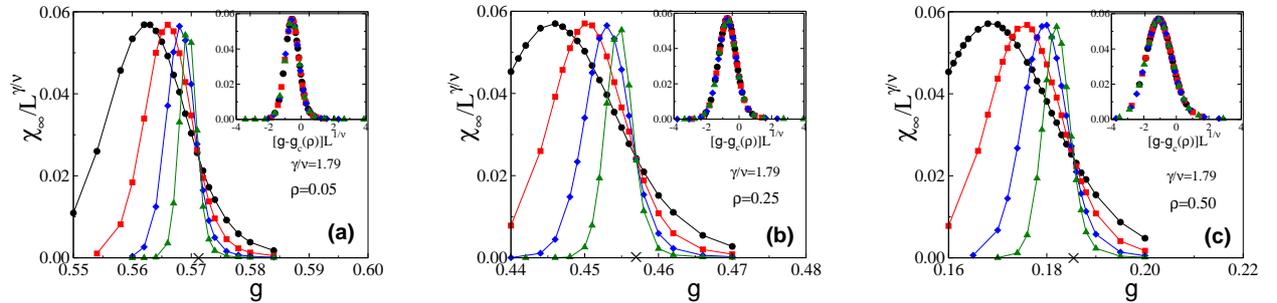

  \centerline{\psfig{file=santra_fig6a.eps,width=0.3\textwidth}
    \hfill\psfig{file=santra_fig6b.eps,width=0.3\textwidth}
  \hfill\psfig{file=santra_fig6c.eps,width=0.3\textwidth}}
\caption{\label{xinf} Plot of $\chi_{\infty}/L^{\gamma/\nu}$ against
  $\mathtt{g}$ for three different $\rho= 0.05$ (a), $0.25$ (b) and
  $0.50$ (c) for the same set of system size as used in
  FIG. \ref{pinf}. The plots of $\chi_{\infty}/L^{\gamma/\nu}$ for
  different $L$ are passing through a single point at
  $\mathtt{g}=\mathtt{g}_c(\rho)$ for the particular value of
  $\gamma/\nu=1.79$. In the inset of figures
  $\chi_{\infty}/L^{\gamma/\nu}$ are plotted against
  $[\mathtt{g}-\mathtt{g}_c(\rho)]L^{1/\nu}$. A good data collapse is
  obtained for the estimated critical exponents $\gamma/\nu=1.79$.}
\end{figure*}

The variation in the fluctuation of order parameter $\chi_{\infty}$ is
studied as a function of growth parameter $\mathtt{g}$ for the
different values of $L$ and $\rho$.  In Fig. \ref{xinf},
$\chi_{\infty}/L^{\gamma/\nu}$ are plotted against the growth
probability $\mathtt{g}$ for different $L$ at $\rho=0.05$ (a),
$\rho=0.25$ (b), $\rho=0.50$ (c) taking $\gamma/\nu=1.79$ as that of
OP. The plots intersect a precise value of $\mathtt{g}$ corresponding
to $\mathtt{g}$ of respective $\rho$. The maximum values of
$\chi_\infty/L^{\gamma/\nu}$ remain independent over the system size
$L$ at all values of $\rho$ which confirms the value of $\gamma/\nu$
already estimated here. The verification of FSS form of $\chi_\infty$
is given in the respective inset of Fig.\ref{xinf} for different
values of $\rho$. In the inset, $\chi_{\infty}/L^{\gamma/\nu}$ is
plotted against the scale variable
$[\mathtt{g}-\mathtt{g}_c(\rho)]L^{1/\nu}$, taking the values of
$\gamma/\nu$ and $1/\nu$ as that of OP. A good data collapse are found
to occur for all values of $\rho$.

\begin{figure*}[h]
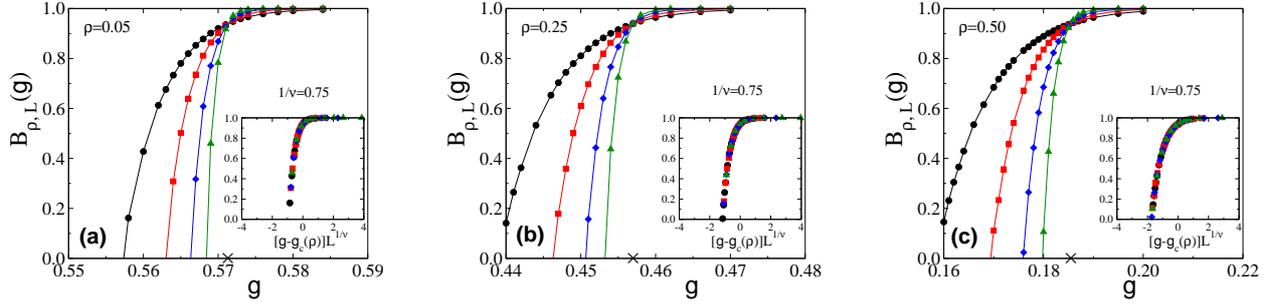

  \centerline{\psfig{file=santra_fig7a.eps,width=0.3\textwidth}
    \hfill\psfig{file=santra_fig7b.eps,width=0.3\textwidth}
  \hfill\psfig{file=santra_fig7c.eps,width=0.3\textwidth}} 
\caption{\label{binder1} Plot of $B_{\rho, L}(\mathtt{g})$ versus
  $\mathtt{g}$ for three different $\rho= 0.05$ (a), $0.25$ (b) and
  $0.50$ (c) for the same set of system size as used in
  FIG. \ref{pinf}. The plot of $B_{\rho, L}(\mathtt{g})$ for different
  $L$ are passing through a single point at thresholds
  $\mathtt{g}=\mathtt{g}_c(\rho)$, marked by cross on
  $\mathtt{g}$-axis. In inset, $B_{\rho, L}(\mathtt{g})$ are plotted
  against $[\mathtt{g}-\mathtt{g}_c(\rho)]L^{\beta/\nu}$. A good data
  collapse is obtained.}
\end{figure*}

\subsection{Binder cumulant}
The values of the critical exponents and the scaling forms of
different geometrical quantities suggest that the PT in TPPM is of
continuous second order transition at all values of $\rho$. In order
to confirm the nature of transition in TPPM, the $4$th order Binder
cumulant (BC) \cite{kbinder},
\begin{equation}
\label{bc}
B_{\rho,L}(\mathtt{g})=\frac{3}{2}\left[1-\frac{\langle
    S_{max}^4\rangle}{3\langle S^2_{max}\rangle^2}\right]
\end{equation} 
is studied. In Fig.\ref{binder1}, $B_{\rho,L}(\mathtt{g})$ is plotted
against $\mathtt{g}$ for different $L$ at $\rho=0.05$ (a), $\rho=0.25$
(b) and $\rho=0.50$ (c). For all values of $\rho$, the plots of
$B_{\rho,L}(\mathtt{g})$ for different $L$ intersect at a point
corresponding to the critical percolation threshold
$\mathtt{g}_{c}(\rho)$ of the respective values of $\rho$ as it occurs
for a continuous PT. The FSS form of BC is given by,
\begin{equation}
\label{bc1}
B_{\rho,L}(\mathtt{g})=
\widetilde{B}[(\mathtt{g}-\mathtt{g}_{c}(\rho))(1-\rho)L^{1/\nu}],
\end{equation}
where $\widetilde{B}$ is a scaling function.  The above scaling form
is verified in the insets of Fig. \ref{binder1}, plotting BC against
$[(\mathtt{g}-\mathtt{g}_{c}(\rho))]L^{1/\nu}$ taking $\nu=4/3$ as that
of OP. Good collapse of data are observed at the respective
$\mathtt{g}_c$ for different values of $\rho$. Thus, for all values of
$\rho$ the model represents second order PT that belongs to the same
universality class of OP.

\begin{figure*}[h]
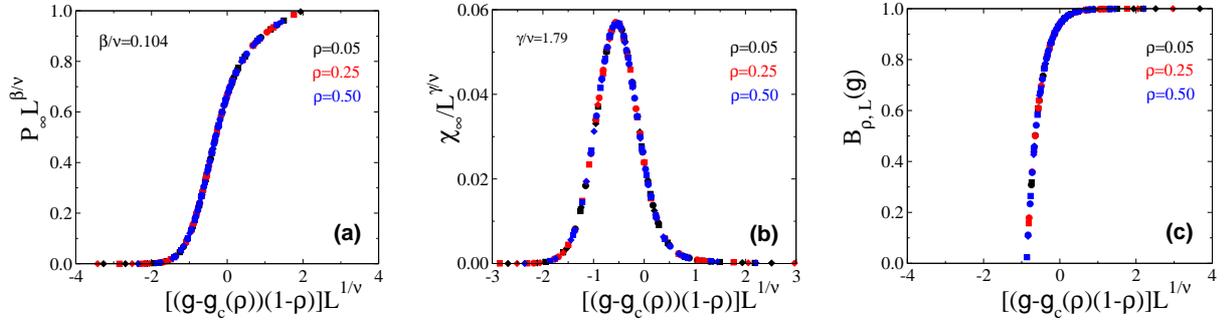

  \centerline{\psfig{file=santra_fig8a.eps,width=0.3\textwidth}
    \hfill\psfig{file=santra_fig8b.eps,width=0.3\textwidth}
    \hfill\psfig{file=santra_fig8c.eps,width=0.3\textwidth}
    \hfill}
\caption{\label{rhosl} Plot of $P_{\infty}L^{\beta/\nu}$ (a),
  $\chi_{\infty}/L^{\gamma/\nu}$ (b), $B_{\rho, L}(\mathtt{g})$ (c)
  and $\chi'^2/L^{\gamma/\nu}$ (d) against scaled variable
  $[(\mathtt{g}-\mathtt{g}_c(\rho))(1-\rho)]L^{1/\nu}$ for three
  different $\rho=0.05$ (black), $0.25$ (red) and $0.50$ (blue) and
  for three different $L=512(\bigcirc)$, $1024(\square)$ and
  $2048(\triangle)$. A good data collapse is obtained for all values
  of $L$ and $\rho$.}
\end{figure*}
\subsection{Scaling with $\rho$}
Finally, we verify the scaling form of the all the above geometrical
quantities as a function of $\rho$, the initial seed concentration. In
Fig.\ref{rhosl}, the respective scaling forms of $P_{\infty}$
(Eq.\ref{fsc2}), $\chi_{\infty}$ (Eq.\ref{chiinffss}), and $B_{\rho,
  L}(\mathtt{g})$ (Eq.\ref{bc1}) are plotted against the scaled
variable $[(\mathtt{g}-\mathtt{g}_c(\rho))(1-\rho)]L^{1/\nu}$ for
three different values of $\rho$. It can be seen that a very good
collapse of data of different $\rho$ occurs for all three geometrical
properties \cite{santraroy}. The scaling form presumed for the cluster
size distribution in Eq.\ref{nsg} is found to be correct. This is
because of the fact that the change in area fraction from its critical
value is just proportional to
$(\mathtt{g}-\mathtt{g}_c(\rho))(1-\rho)$ in the critical regime.

\section{Conclusion}
A new two parameter percolation model with multiple cluster growth is
developed and studied extensively following finite size scaling
hypothesis. In this model, two tunable parameters are the initial seed
concentration $\rho$ and the cluster growth probability
$\mathtt{g}$. It is found that for each $\rho$ there exists a critical
growth probability $\mathtt{g}_c$ at which a continuous percolation
transition occurs. A finite size scaling theory for such percolation
transition involving $\rho$ and $\mathtt{g}$ is proposed and verified
numerically. It is found that the values of the critical exponents
describing the scaling functions at the criticality in this model are
that of ordinary percolation for all values of $\rho$. Hence, all such
transitions belong to the same universality class of percolation. A
phase line consisting of second order phase transition points is found
to separate the connected region from the disconnected region in the
$\rho-\mathtt{g}$ parameter space. No first order transition is found
to occur at any $\rho$ as there is no suppression in the growth of
specific clusters.


\section{References}
\begin{thebibliography}{10}
\bibitem{flory} P. J. Flory, J. Am. Chem. Soc. {\bf 63}, 3083, 3091,
  3096 (1941).
\bibitem{broad} S. R. Broadbent and J. M. Hammersley, Percolation
  processes I. Crystals and mazes, Proc. Camb. Philos. Soc. {\bf 53},
  629, 641 (1957).
\bibitem{king} P. R. King {\em et al.}, Physica A {\bf 274}, 60
  (1999); Physica A {\bf 314}, 103 (2002)
  
\bibitem{cardy} J. L. Cardy and P. Grassberger, J. Phys. A:
  Math. Gen. {\bf 18}, L267 (1985).

\bibitem{cohen} R. Cohen, D. Ben-Avraham and S. Havlin, Phys. Rev. E
  {\bf 66}, 036113 (2002).

\bibitem{Acin} A. Acin, J. I. Cirac and M. Lewenstein, {\em Nature Physics}
  {\bf 3}, 256 (2007).

\bibitem{roux} H. J. Herrmann and S. Roux, editors, {\em Statistical Models
  for the Fracture of Disordered Media}, North-Holland, 1990.
  
\bibitem{ball} Z. Ball, H. M. Phillips, D. L. Callahan and
  R. Sauerbrey, Phys. Rev. Lett. {\bf 73}, 2099 (1994).
  
\bibitem{roman} H. E. Roman, A. Bunde and W. Dieterich, Phys. Rev. B
  {\bf 34}, 3439 (1986).
  
\bibitem{sahimi} M. Sahimi, {\em Applications of Percolation Theory},
  Taylor and Francis, London, 1994.

\bibitem{stauffer} D. Stauffer and A. Aharony, {\em Introduction to
  Percolation Theory}, second edition, Taylor and Francis, London,
  Washington, DC, 1992.

\bibitem{ziffp} M. E. J. Newman and R. M. Ziff, Phys. Rev. Lett. {\bf
  85}, 4104 (2000).
  
\bibitem{bsp} J. Chalupa, P. L. Leath, G. R. Reich, J. Phys. C, Solid
  State Phys. {\bf 12}, L31–L35 (1979).

\bibitem{dp} S. P. Obukhov, Physica A {\bf 101}, 145 (1980);
  H. Hinrichsen, Brazilian Journal of Physics {\bf 30}, 69 (2000).

\bibitem{sp} S. B. Santra and I. Bose, J. Phys. A {\bf 24}, 2367
  (1991); S. B. Santra and I. Bose, J. Phys. A {\bf 25}, 1105 (1992).

\bibitem{dpsp} S. B. Santra, Eur. Phys. J. B. {\bf 33}, 75 (2003);
  S. Sinha and S. B. Santra, Eur. Phys. J. B. {\bf 39}, 513 (2004).

\bibitem{bunde} A. Aharony, in {\em Fractals and Disordered systems}
  edited by A. Bunde and S. Havlin, Springer-Verlag, Berlin, (1991).
  
\bibitem{epj223} N. Ara\'ujo, P. Grassberger, B. Kahng, K. J. Schrenk,
  and R. M. Ziff, Eur. Phys. J. Special Topics {\bf 223}, 2307 (2014)
  and references therein.
  
\bibitem{saberi} A. A. Saberi, Phys. Rep. {\bf 578}, 1 ( 2015 ).

\bibitem{AP} D. Achlioptas, R. M. D'Souza, and J. Spencer, {\em Science}
{\bf 323}, 1453 (2009). 

\bibitem{cep} P. Grassberger, C. Christensen, G. Bizhani, S.-W. Son, and
M. Paczuski, Phys. Rev. Lett. {\bf 106}, 225701 (2011).

\bibitem{janssen2016} H.-K. Janssen and O Stenull, EPL, {\bf 113},
  26005 (2016).

\bibitem{HK} J. Hoshen and R. Kopelman, {\em Phys. Rev. B} {\bf 14}, 8
  (1976). 

\bibitem{LH} P.L. Leath, {\em Phys. Rev. B} {\bf 14}, 5046  (1976).
 
\bibitem{santraroy} B. Roy and S. B. Santra, Croat. Chem. Acta. {\bf
  86}, 495 (2013).

\bibitem{arg} N. Tsakiris, M. Maragakis, K. Kosmidis, and
  P. Argyrakis, {\em Phys. Rev. E} {\bf 82}, 041108 (2010), {\em
    Eur. Phys. J.B} {\bf 81}, 303 (2011).

\bibitem{binder} K. Binder, Z. Phys. B {\bf 43}, 119 (1981).
  
\bibitem{bruce} A.D. Bruce, J. Phys. C {\bf 14}, 3667 (1981).
  
\bibitem{kbinder} K. Binder, Rep. Prog. Phys. {\bf 60}, 487 (1997).

\end{thebibliography}
\end{document}